\pdfoutput=1
\documentclass[
10pt,
aip,
amsmath,
amssymb,
twocolumn,
letterpaper,
nobalancelastpage,
final,
citeautoscript,
floatfix,
raggedbottom,
superscriptaddress
]{revtex4-1}

\usepackage{newtxtext,newtxmath}
\usepackage[scaled=0.875]{helvet}
\usepackage[usenames,dvipsnames]{color}
\usepackage{graphicx}
\usepackage{microtype}
\usepackage[bookmarks=false,colorlinks]{hyperref}
\usepackage{xfrac}
\hypersetup{
    linkcolor=RubineRed,        
    citecolor=RoyalBlue,      
    filecolor=Mulberry,     	
    urlcolor=RoyalBlue          
}

\usepackage{ccaption}
\usepackage{ragged2e}
  \captionnamefont{\small \sffamily \bfseries }
  \captiontitlefont{\small }
  \captiondelim{.~}
  \captionstyle{\justifying}



\newcommand{\RN}[1]{%
  \textup{\uppercase\expandafter{\romannumeral#1}}
}

\newcommand{\ai}[1]{\textcolor{black}{#1}}
\newcommand{\rw}[1]{\textcolor{black}{#1}}

\begin{document}

\title{\Large Featureless adaptive optimization accelerates functional electronic materials design}

\author{\large Yiqun Wang}
\affiliation{Department of Materials Science and Engineering, Northwestern University, Illinois 60208, USA}

\author{\large Akshay Iyer}
\affiliation{Department of Mechanical Engineering, Northwestern University, Illinois 60208, USA}

\author{\large Wei Chen}
\affiliation{Department of Mechanical Engineering, Northwestern University, Illinois 60208, USA}

\author{\large James M.\ Rondinelli}
\email{jrondinelli@northwestern.edu}
\affiliation{Department of Materials Science and Engineering, Northwestern University, Illinois 60208, USA}


\begin{abstract}
Electronic materials exhibiting phase transitions between metastable states (e.g.,  metal-insulator transition materials with abrupt electrical resistivity transformations) are challenging to decode. For these materials, conventional machine learning methods display limited predictive capability due to data scarcity and the absence of features impeding model training. In this article, we demonstrate a discovery strategy based on multi-objective Bayesian optimization to directly circumvent these bottlenecks by utilizing latent variable Gaussian processes combined with high-fidelity electronic structure calculations for validation in the chalcogenide lacunar spinel family. We directly and simultaneously learn phase stability and band gap tunability from chemical composition alone to efficiently discover all superior compositions on the design Pareto front. Previously unidentified electronic transitions also emerge from our featureless adaptive optimization engine. Our methodology readily generalizes to optimization of multiple properties, enabling co-design of complex multifunctional materials, especially where prior data is sparse.\par
\end{abstract}

\maketitle  

\begin{figure*}
  \centering
  \vspace{-1mm}
 \includegraphics[width=0.98\textwidth]{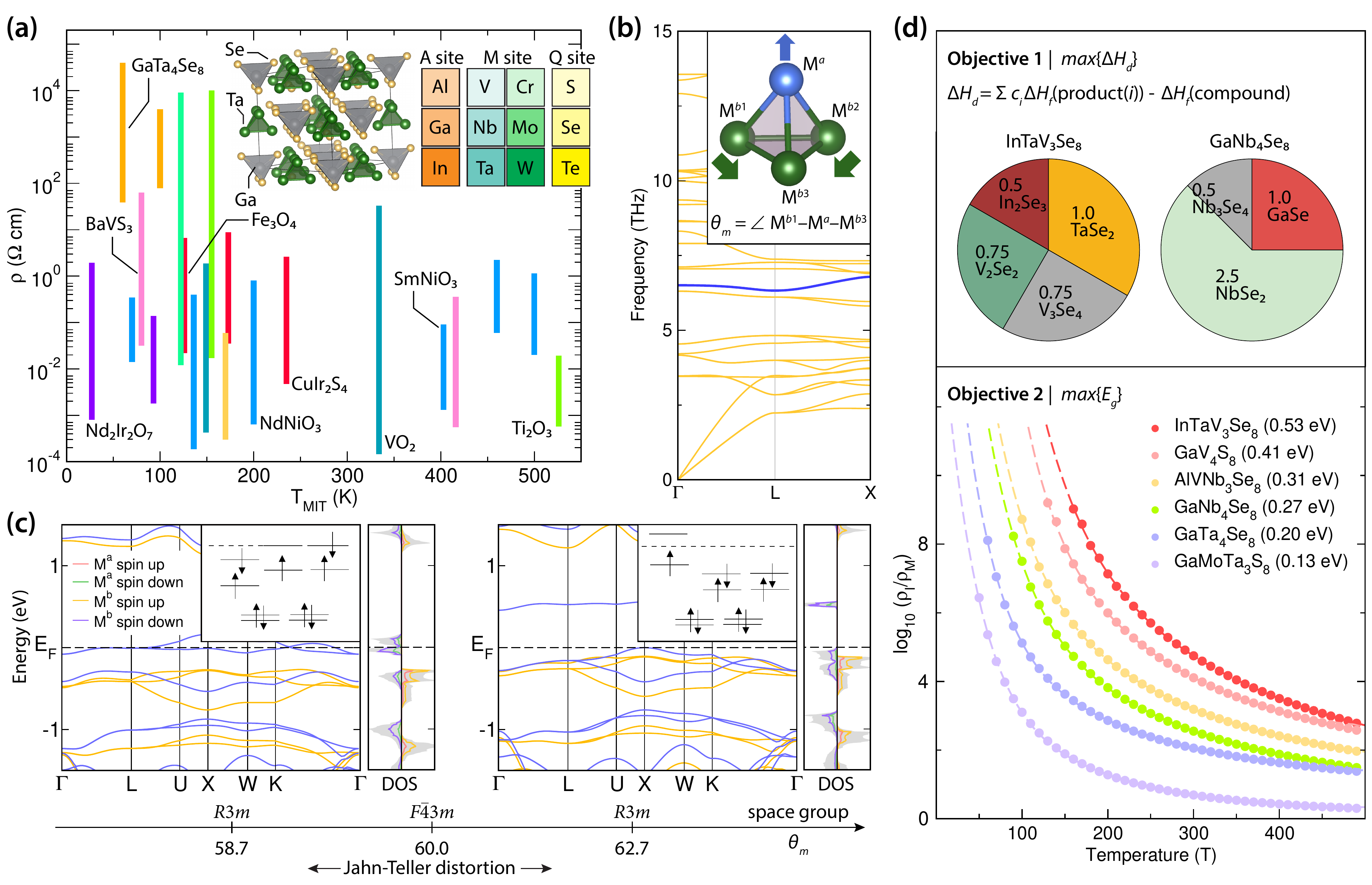}\vspace{-5pt}
    \caption{\sffamily 
    \textbf{Metal-insulator transition materials and design objectives for the lacunar spinel family.} 
    (a) The range in resistivity accessible (length of bar) across the MIT and transition temperature for a variety of MIT materials.
    (left inset) The crystal structure of GaTa$_4$Se$_8$.
    (right inset) Candidate elements on each site of the lacunar spinel  structure. 
    (b) DFT-simulated phonon dispersion curves of GaMo$_4$S$_8$ in the rhombohedral ground state, 
    the blue curve corresponds to the Jahn-Teller active cluster distortion mode. 
    (inset) The transition-metal cluster with a single apical M$^a$ atom and three basal M$^b$ atoms. 
    The arrows indicate displacements characterizing 
    the Jahn-Teller active phonon mode. 
    The intra-tetrahedral cluster angle $\theta_m$ formed by  M$^{b1}$-M$^a$-M$^{b2}$.
    (c) Electronic band structures and projected density of states (DOS in units of states/eV/spin/f.u.) of GaMo$_4$S$_8$ in its (right) semiconducting ground state  and (left) metallic metastable phase with $\theta_m$. %
    The two $R3m$ phases are connected by the Jahn-Teller-type structural distortion 
    with a $F\bar{4}3m$ intermediate state. 
    (insets) Molecular orbital diagrams of the Mo$_4$ cluster with different local geometries. 
    (d) Design Objective 1 with the definition of decomposition enthalpy change and the graphical decomposition pathways of two lacunar spinels for demonstration.
    The DFT-simulated temperature-dependent log ratio of the resistivity in the insulating and metallic phases of lacunar spinels, including experimentally known compounds and newly discovered compositions, serves as design Objective 2. DFT band gaps specified in parentheses.}
  \label{fig:fig1}
\end{figure*}

\section{Introduction}

Upon traversing a critical temperature, the electrical resistivity of 
a metal-insulator transition (MIT) material can change by orders of magnitude\cite{Imada1998}.
Athermal approaches may also trigger the electronic transitions, 
including (chemical) pressure, variable carrier-densities, and applied electromagnetic fields.
The transformations can be used to  
encode, store, and process information for 
beyond von-Neumann microelectronics and overcome performance 
limits of conventional field-effect transistors\cite{Shukla2015} for advanced 
logic/memory technologies.\cite{YouZhou2015}
Because macroscopic MITs occur in materials with diverse chemistries and structures (\autoref{fig:fig1} (a)), 
various microscopic mechanisms -- electron-lattice interactions, electron-electron 
interactions, or a combination thereof -- lead to large variations in critical temperatures and accessible resistivity changes\cite{Yang2011,Zhang2019b}. 
This diversity exacerbates the efficient discovery and 
optimization challenge of achieving multiple property requirements 
to outperform silicon-based devices,\cite{Coll2019}  
including stability, 
large reversible resistivity changes ($\approx\!\!10^5$), 
and above room-temperature operation.
The aforementioned complexity is ubiquitous in formulating atomic scale material chemistry and macroscopic functionality relationships to guide property optimization. Presently, the  
principal solution relies on a better understanding of the underlying materials physics.
%
Numerous data-driven machine learning models, however, have shown promising results in deciphering nonlinear relationships 
between materials structure and properties
when sufficient training data is available
\cite{Butler2018,Agrawal2016,xie2018crystal,Schmidt2019,noh2019inverse}. 
The predictive performance (error and efficiency) of these approaches is limited by the quality and quantity of the data, typically $>\mathcal{O}(10^2)$, which poses a severe challenge to MIT materials design owing to the relatively small size of available dataset of $\approx\mathcal{O}(10^1)$. \ai{The suitability of the machine learning model is determined by the input dimensionality and dataset size, which for high dimensional inputs necessitates large datasets and complex models for good predictive performance.}
%
A number of sequential materials design strategies have recently emerged\cite{ling2017high,seko2015prediction,lookman2019active, gopakumar2018multi} to rescue the lack of data problem. 
Mostly being based on the Bayesian approach, 
these methods utilize knowledge extracted from existing data to infer properties of unknown materials following a step-by-step discovery manner. 
This sequential optimization method fits well with the regular materials discovery procedure both experimentally and computationally, 
since property evaluations are usually time and effort consuming (e.g. synthesis and simulations).
Nevertheless, these sequential learning models 
typically rely on numerical materials descriptors (features)
whose selection may be informed 
by domain knowledge or trial-and-error approaches. 
For MIT materials systems which lack of microscopic
understanding in how different compositions influence the phase transitions,
this leads to ambiguity in feature formulation for discovery of MIT materials from structure and composition alone rather than through effective Hamiltonians\cite{Zhang2019b}.

What could we do when there is little data available while the governing materials physics is not abundantly clear? 
Here we demonstrate a generic strategy to 
overcome the data scarcity as well as the feature engineering problems. 
We utilize multi-objective Bayesian optimization (MOBO) 
with latent-variable Gaussian processes (LVGP) 
to simultaneously optimize the band gap tunability and \rw{thermal stability}
in a family of candidate MIT materials -- the lacunar spinels (introduced in the next section). 
%
\rw{With the goal to identify the optimal compositions, among hundreds of possible chemical combinatorics with both high functionality as well as synthesizability, 
we successfully retrieved all 12 superior compositions on the Pareto front}
by searching through a small fraction of the total design space.
Notably, the chemical compositions (i.e., element on each crystallographic site) are all the model requires to guide this discovery procedure.
No handcrafted features are required in this method, hence featureless learning, 
making our methodology easily generalizable to other materials design problems. 
We also showcase how this model could offer helpful guidance on making better decisions towards the optimal design---selecting the next candidate compound to synthesize or simulate.
Our adaptive optimization engine (AOE) frees  researchers from exclusively relying 
on their chemical intuition, which can require an entire career to accumulate, and is particularly valuable when the research budget is limited.

\begin{figure*}
  \centering
  \vspace{-1mm}
 \includegraphics[width=0.95\linewidth]{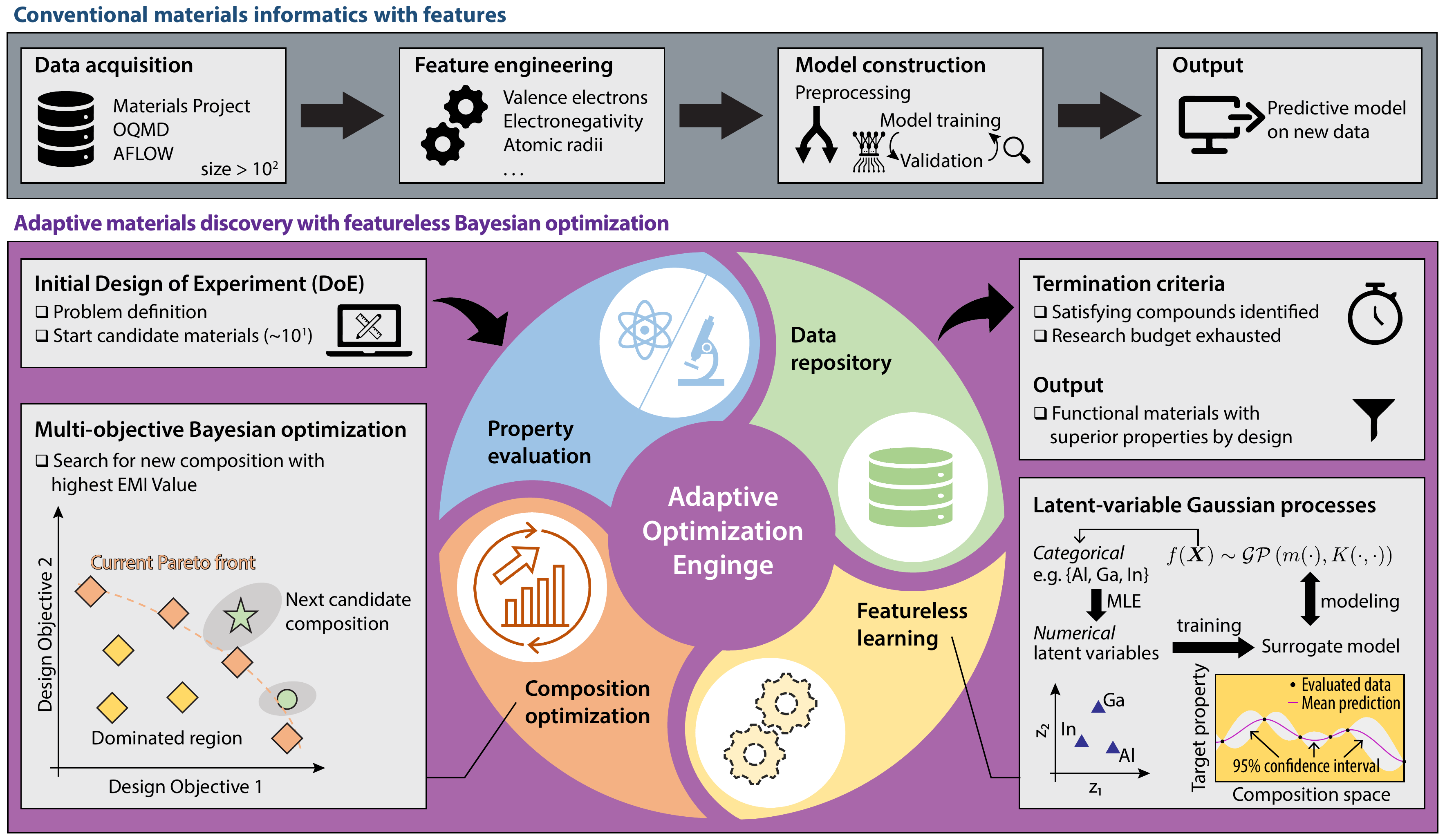}\vspace{-2pt}
  \caption{\sffamily  
  \textbf{Comparison of conventional (feature-required) machine learning with the   featureless adaptive optimization engine.}
  \textbf{Upper panel}, 
  The workflow of a conventional feature-based machine learning model typically involves data acquisition, feature engineering, model construction, and property prediction.
  \textbf{Lower panel}, The adaptive materials discovery scheme starts from an initial set of design of experiments (DoE),
  where system variables, design objectives, and design space are first defined for the problem,  
  providing a few $\mathcal{O}(10^1)$ candidate materials to initialize the 
  discovery procedure.
  \textbf{Property evaluation}, The target material properties (design objectives) are evaluated either by experimental measurement or theoretical simulations.
  Candidate composition and its evaluated properties 
  are then added to a \textbf{Data repository}, which initially may 
  either be empty or only contains entries for existing materials within the design space.
  Its size grows as more candidate materials are evaluated during the adaptive optimization process.
  \textbf{Featureless learning} involves
  directly learning from the chemical composition of materials comprising the data repository by mapping 
  each compositional variable 
  into a two-dimensional latent space (spanned by $z_1$ and $z_2$) using maximum likelihood estimation, 
  which enables the construction of a latent variable Gaussian process (LVGP) surrogate model.
  One surrogate model is constructed for each design objective using all currently available knowledge within the data repository.
  \textbf{Composition optimization}, 
  Multi-objective Bayesian optimization is then performed with the LVGP models to obtain the next candidate material composition with the highest expected maximin improvement (EMI) value.
  The model accounts for uncertainty with the 
  $95\%$ confidence interval shown as the shadowed area around the new compositions (the green symbols).
  In the lower left inset, the green star composition outperforms the green circle composition, and will be passed to the next property evaluation procedure.
  The iterative optimization step continues until all compounds satisfying the objectives are discovered, forming the Pareto front, or computational resources expire. 
   \label{fig:fig2}}
\end{figure*}

\section{Materials Design Objectives}

The complex lacunar spinel family $A$M$^a$M$^b_3Q_8$ with trivalent main group $A$, transition metal M, 
and chalcogenide $Q$ ions demonstrate the complexity active in MIT materials design.
The structure comprises 
transition-metal clusters (TMC)  with 
M$^a$ and M$^b$ cations at the apical and basal positions of the tetrahedra 
(\autoref{fig:fig1} (b) inset).
Although there are hundreds of possible elemental combinations on the four lattice sites in the crystal structure (\autoref{fig:fig1} (a)), only tens of the lacunar spinels have been experimentally reported. \cite{cario2010electric, powell2007cation}
%
For example, GaV$_4$S$_8$ ($\mathrm{M}^a\!=\!\mathrm{M}^b\!=\!\mathrm{V}$) exhibits a MIT\cite{camjayi2014first},
exotic spin textures\cite{kezsmarki2015neel}, 
and multiferroism\cite{wang2015polar}
while GaVTi$_3$S$_8$ shows negative magnetoresistance and half-metallic ferromagnetism.\cite{Dorolti2010} 
%
Most lacunar spinels are narrow-bandwidth semiconductors in their ground states\cite{pocha2000electronic,cario2010electric};
these electronic properties are governed by distortions of the local TMC from the ideal $T_d$ geometry,\cite{sieberer2007importance} 
which manifest as low-frequency phonons as shown for GaMo$_4$S$_8$
(\autoref{fig:fig1} (b), blue curve).
Jahn-Teller-type distortions, which correspond to elongation along the [111] direction alter the TMC geometry, are particularly important; they transform the insulating GaMo$_4$S$_8$ ground state into a metastable metallic phase 
(\autoref{fig:fig1} (c)).
%
The MIT arises from a redistribution of electrons among the structure-driven orbital hierarchy (\autoref{fig:fig1} (c) insets).
Furthermore, these phases host low energy electronic structures, discernible from the 
projected density of states (pDOS) in \autoref{fig:fig1} (c), 
that arise from the different 
M$^a$ and M$^b$ sites. 
%
This capability to exhibit
distinct and tunable 
electronic phases poses a challenge in the design of lacunar spinels from physics-based models while also making them an ideal system for MIT performance optimization.


In pursuit of novel MIT materials with superior performance, we specifically
seek lacunar spinels that exhibit high thermodynamic stabilities and large resistivity-switching ratios, 
which we formulate as two design objectives for our materials discovery task.
%
We reduce the approximately 
$\mathcal{O}(10^3)$ compositional space to 
270 candidates that maintain a 
1\,M$^a$ to 3\,M$^b$ ratio. 
($A$M$^a_2$M$^b_2Q_8$ compositions
are excluded as they remove the $C_{3v}$ symmetry fundamental to the MIT; 
Cr is also excluded from occupying the M$^b$ site, because it destabilizes\cite{bichler2007tuning} the cluster.)
\rw{This design space extends the known composition space that have been experimentally synthesized; therefore, it is important to determine the crystal stability, i.e., whether the selected chemical combination forms a thermodynamically stable lacunar spinel structure.}
\rw{To that end, we} define the first design objective as the decomposition enthalpy change ($\Delta H_d$, \autoref{fig:fig1} (d)), 
and use density functional theory (DFT) simulations to evaluate formation energies (see Appendix A).
%
Materials with larger $\Delta H_d$ are expected to be more synthesizable\cite{Bartel2018} and stable during operation, 
making it a useful filter to prioritize compounds for subsequent theoretical analysis and 
synthetic processing.
The second design objective is the ground state band gap ($E_g$).
We use it as a proxy for the resistivity-switching ratio 
since $E_g$ is positively correlated with the resistivity change between different electronic states (\autoref{fig:fig1} (d)). 
A larger $E_g$ also allows for greater band-gap tunability through control over the $C_{3v}$ distortion, 
which is a desirable 
feature for programmable electronics.
Importantly, because $E_g$  is small for most MIT materials,
stability is expected to be lower and more difficult to achieve than that of 
nonpolymorphous compounds with majority ionic or covalent bonding.\cite{Burdett1988} 
%

\section{Adaptive Optimization Engine (AOE)}
The nonlinear responses of 
both design objectives bring severe challenges to compound optimization 
beyond those amplified by chemical combinatorics using data-driven models.
We overcome these obstacles by implementing 
a cyclic adaptive optimization engine
shown in \autoref{fig:fig2}, 
which consists of four iterative tasks (\emph{vide infra}): 
property evaluation, aggregation of data (in a repository), featureless learning, and composition optimization.
Beyond returning a predictive model capable of predicting properties from compositions alone, 
our iterative AOE leverages earlier approaches\cite{ling2017high,seko2015prediction,lookman2019active} to deliver materials with superior performance by design of composition-based solutions. 
%
In contrast to single objective design which often has a unique solution, multiobjective design aims to uncover the Pareto front---a set of non-dominated designs where no individual objective can be improved without deterioration in other objectives. In other words, the Pareto front represents the optimal trade-offs that can be achieved amongst competing objectives. 
There is no relative importance of multiple objectives in the process of identifying the Pareto front, which simply offers the designer several options from which to select the subset of compositions for further investigation and  development. Since the designer's preference may be subjective or informed by other criteria (e.g. cost), herein we present only the  framework for Pareto front discovery and its comprising compositions.
%

The AOE has the important advantage of bypassing the feature engineering procedure as in conventional ML methods; 
it learns properties directly from the chemical composition at each site (i.e., $A$, M$^a$, M$^b$, $Q$).
\ai{Gaussian Process (GP) is ideally suited for this problem, because (a) it interpolates data and hence is ideal for surrogating deterministic responses such as DFT results, and (b) it provides a principled statistical representation for uncertainty quantification, which is essential for Bayesian optimization. Latent-variable methods provide a fundamentally different approach to modelling categorical design variables by alleviating the need for handcrafted features (see Appendix B).
It transforms categorical variables (i.e., elemental compositions)
into a continuous numerical space.
Utilizing these approaches in the AOE, we achieve featureless learning  
and then perform composition optimization under the multiple objectives 
through latent variable Gaussian processes (LVGP).}
%

\begin{figure}
  \centering
  \vspace{-1mm}
 \includegraphics[width=0.98\linewidth]{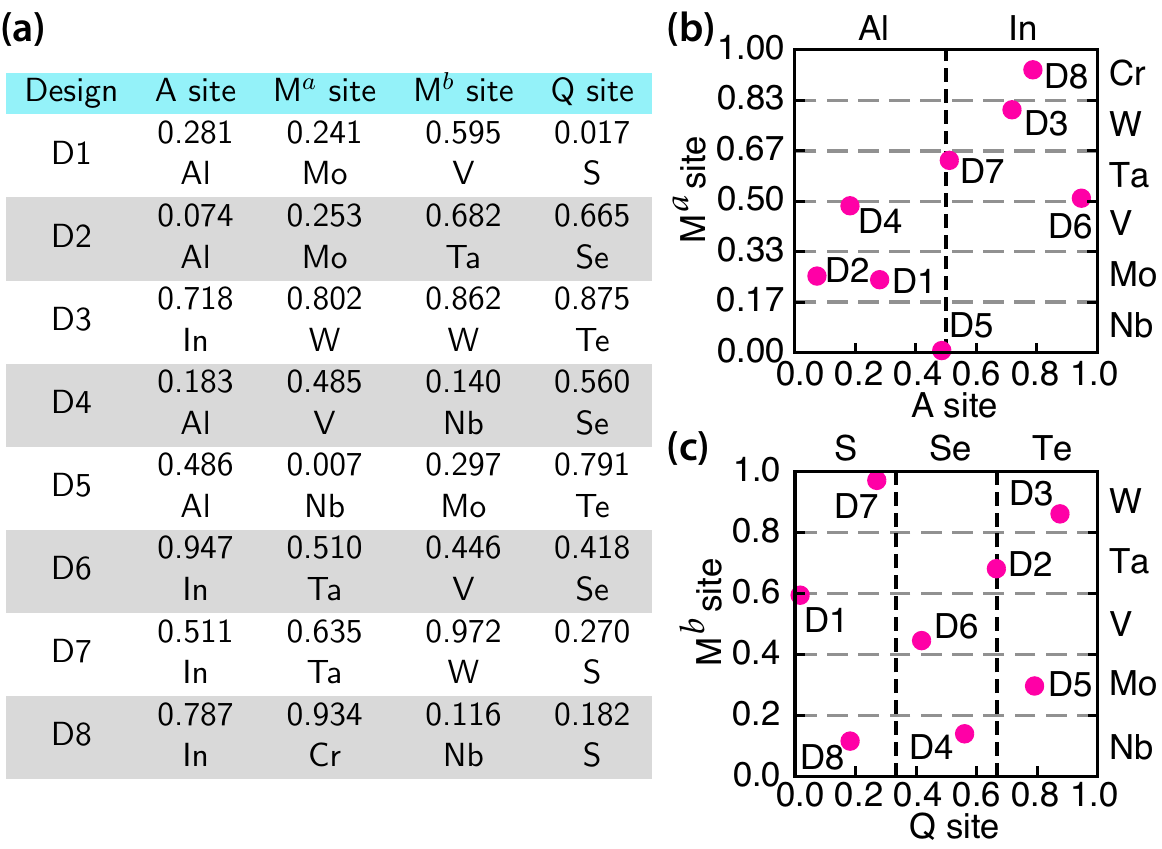}\vspace{-4pt}
 \caption{\sffamily\textbf{Design of experiment (DoE) for the complex lacunar spinel family.}  
(a) A four-dimensional Latin Hypercube Design of size eight is generated,
where each dimension corresponds to a crystal site (e.g. $A$, M$^a$, etc.).
Since the four known compounds are all gallium-based, 
we only consider Al and In for the $A$ site design. 
(b, c) Each dimension is evenly divided into a number of grids, 
each grid represents one candidate elemental composition at that crystal site. 
For instance, the $Q$ site is divided into three grids 
because there are three candidate elements (S, Se, Te) on that site.
The designed composition could then be determined using the grid-composition correspondence.
For example, Design ID Number 1 (D1) resides in the grid corresponding to \{Al, Mo, V, S\}; therefore, its composition is  AlMoV$_3$S$_8$.}
 \label{fig:fig3}
\end{figure}

We start the MIT-materials AOE for the lacunar spinel family 
through an initial design of experiment (DoE) 
consisting of four experimentally known compounds within the family (i.e., GaMo$_4$S$_8$, GaV$_4$S$_8$, GaNb$_4$Se$_8$, and GaTa$_4$Se$_8$)
and eight new compositions generated by 
discretized Latin Hypercube Design (LHD)\cite{mckay1979comparison} (\autoref{fig:fig3}). 
This procedure ensures a variety of elemental combinations within the initial DoE set, 
where each candidate element will appear at least once,
so that the model has knowledge about different elemental contributions to the design objectives.

\begin{figure*} 
  \centering
  \vspace{0mm}
 \includegraphics[width=0.98\linewidth]{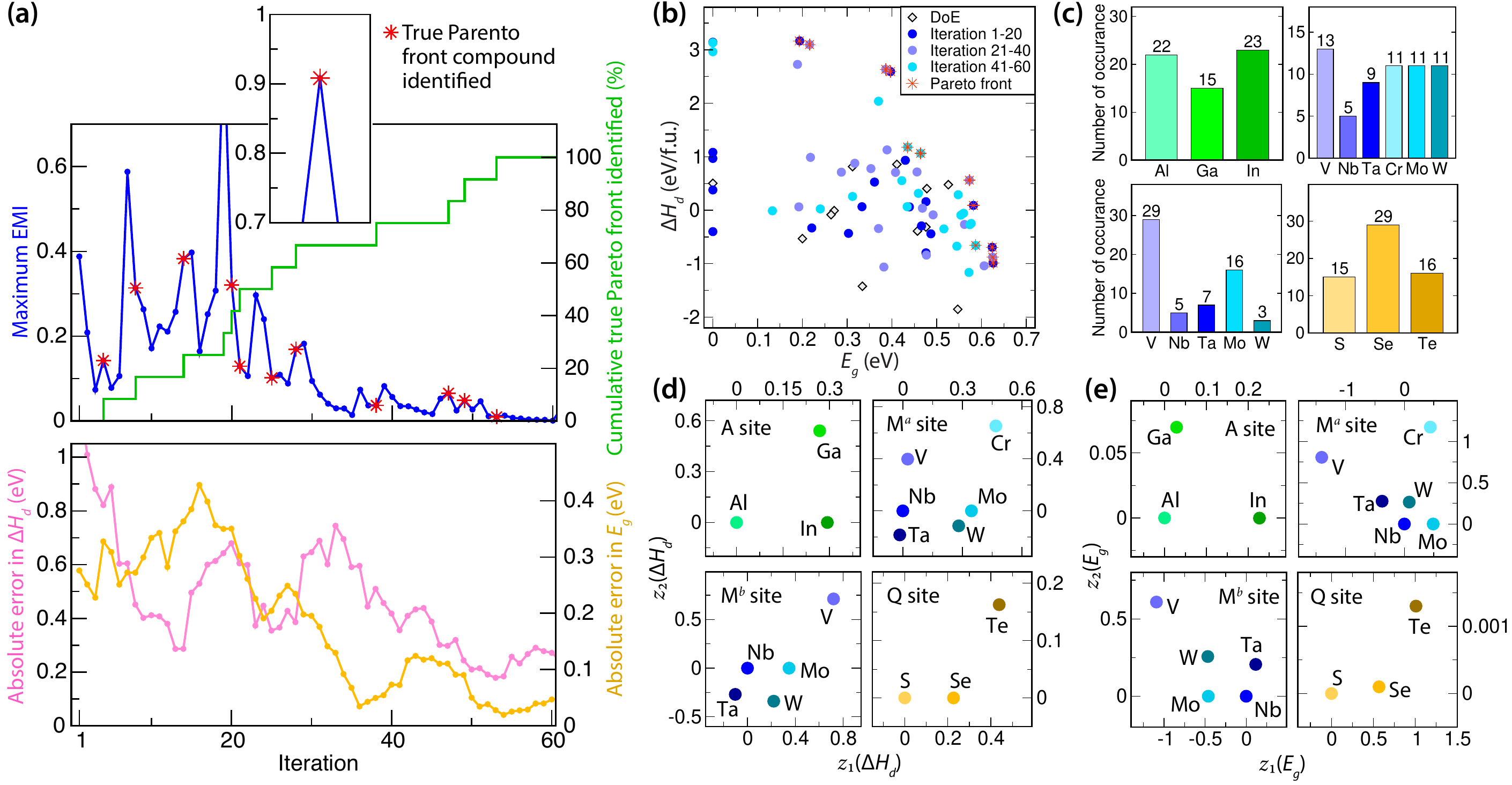}\vspace{-4pt}
  \caption{\sffamily 
  \textbf{The results of adaptive optimization on the lacunar spinel family.}
  (a), Upper panel: Evolution of the highest expected maximin improvement (EMI, blue line) and percentage of true Pareto front compounds identified (green line) as a function of iteration number.
  Results of the first 60 iterations are shown here.
  The red \rw{asterisks} represent sampling points where a true Pareto front design is successfully identified.
  Lower panel: The moving average of 
  absolute error in the predicted $E_g$ and $\Delta H_d$ values for a compound selected by the acquisition function for property evaluation. 
  (b), The distribution of initial design of experiment and the first 60 evaluated compounds.
  Compounds evaluated in earlier stages have darker colors.
  True Pareto front designs are marked with red stars.
  (c), Distribution of Bayesian optimization-sampled elemental compositions for the first 60 iterations. 
  (d, e), Latent space representation of elemental composition at different crystal structure sites in the $\Delta H_d$ and $E_g$ surrogate model, respectively. Results obtained after 60 iterations.
  }
  \label{fig:fig4}
\end{figure*}

Next, we use high-fidelity DFT simulations to evaluate $\Delta H_d$ and $E_g$ (see Appendix A).
This is the most resource-intensive step among the four tasks; 
therefore, it is desirable to iterate 
through the AOE (property evaluation) step 
as few times as possible. Although it is application dependent, AOE can be terminated if a compound with target properties is discovered or the budget (computational/experimental) has been exhausted.
Then, we create a data repository that contains entries for both composition and the evaluated properties. 
Unlike other ML methods, we do not rely on a large number of existing data at either the onset or later in the learning process.
%

We then construct a LVGP model by mapping the
elemental compositions (e.g., Al, Ga, In) 
into a two-dimensional (2D) latent space (\autoref{fig:fig2}, lower right inset) 
where the relative positions of elements are obtained using maximum likelihood estimation (MLE).
This latent space representation enables us to construct
Gaussian process surrogate models for the unknown underlying design objectives, 
$\Delta H_d$ and $E_g$, as a function of composition.
The MOBO 
step then begins and we use the LVGP models to  predict 
$\Delta H_d$ and $E_g$ of the \emph{unexplored} compositions in our design space; 
we choose the next candidate composition for evaluation using 
the expected maximin improvement (EMI, see Appendix B) 
as the acquisition function, 
which quantitatively describes the performance gain compared against the compositions at the current Pareto front.
\rw{The EMI is defined in such a way that both objectives have equal weighting, and the objective properties are normalized with respect to the current min-max values (see Appendix B for details).}
This acquisition function considers both exploration of compositions with high uncertainty 
(\autoref{fig:fig2}, shaded ellipses, lower left inset) 
as well as exploitation of candidates with high performance gain.
The composition with highest EMI is then selected for DFT simulation (property evaluation), 
at which point another AOE cycle commences.
%

The aforementioned iterative optimization procedure progresses and explores the available design space.
One new lacunar spinel composition is evaluated and added to repository after each AOE iteration. 
The LVGP models are also updated in each iteration as more knowledge becomes available.
Owing to the high computational cost of the property evaluation process, 
we terminate the optimization process after 
searching through 1/3 of the entire design space.  
%
In order to validate the effectiveness of this method, 
we ultimately evaluated $\Delta H_d$ and $E_g$ with DFT calculations of all 270 compositions 
within the design space by expending approximately $3\times10^6$ CPU hours.

\begin{figure}[t]
  \centering
  \vspace{-1mm}
 \includegraphics[width=0.99\linewidth]{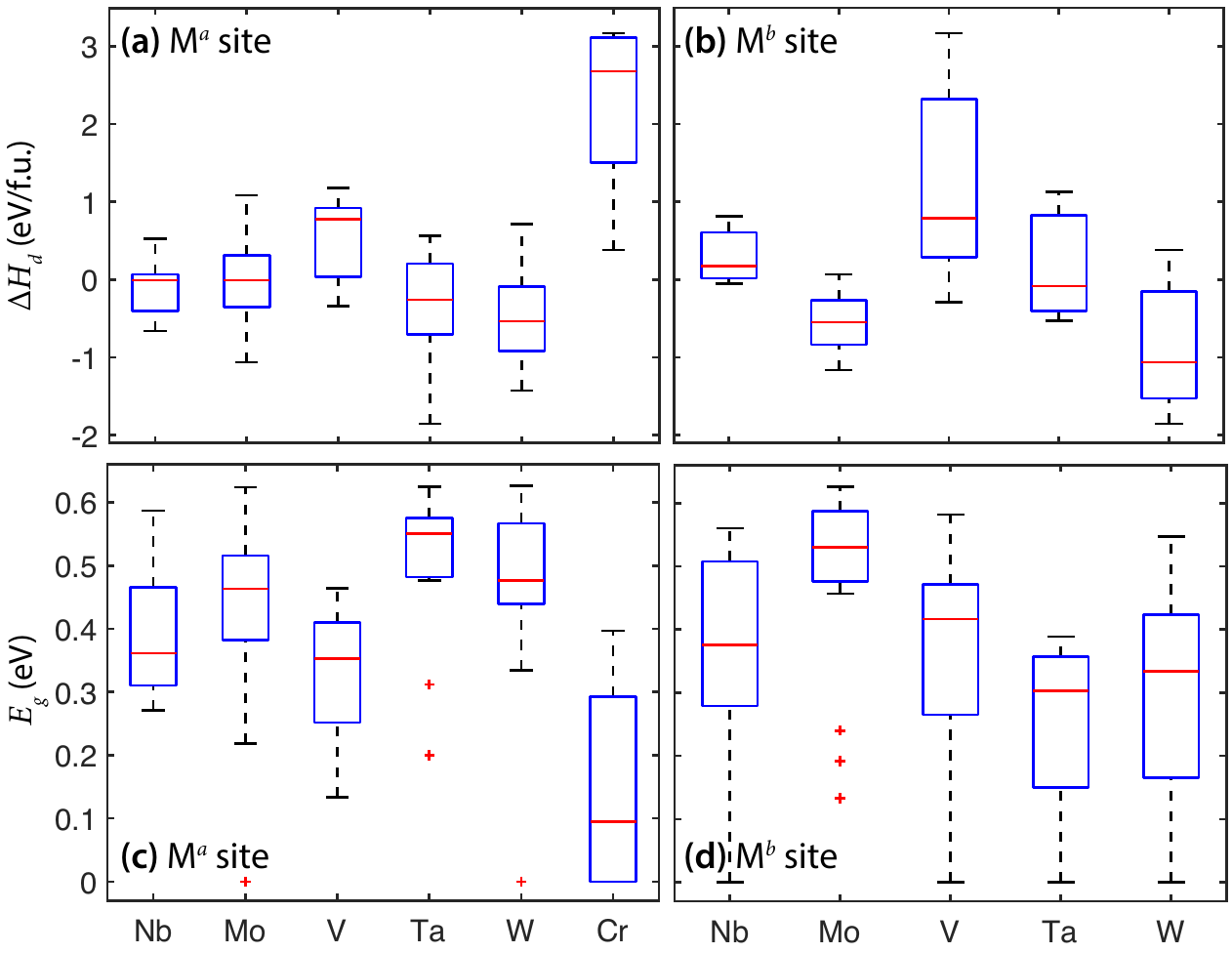}\vspace{-8pt}
  \caption{\sffamily
  \textbf{Composition-property relationships at the transition-metal sites.} Distribution of DFT-evaluated properties of the complex lacunar spinel family with 12 initial DoE sets and 60 iterations of AOE. 
  This data presents the impact different elemental compositions at the transition-metal sites (i.e., M$^a$ and M$^b$) have on the two design objectives (i.e., $\Delta H_d$ and $E_g$). 
  (a, b) decomposition enthalpy change distribution at M$^a$, M$^b$ site. 
  (c, d) band gap distribution at M$^a$, M$^b$ site.}
  \label{fig:fig5}
\end{figure}

\begin{figure}
  \centering
  \vspace{-1mm}
 \includegraphics[width=0.99\linewidth]{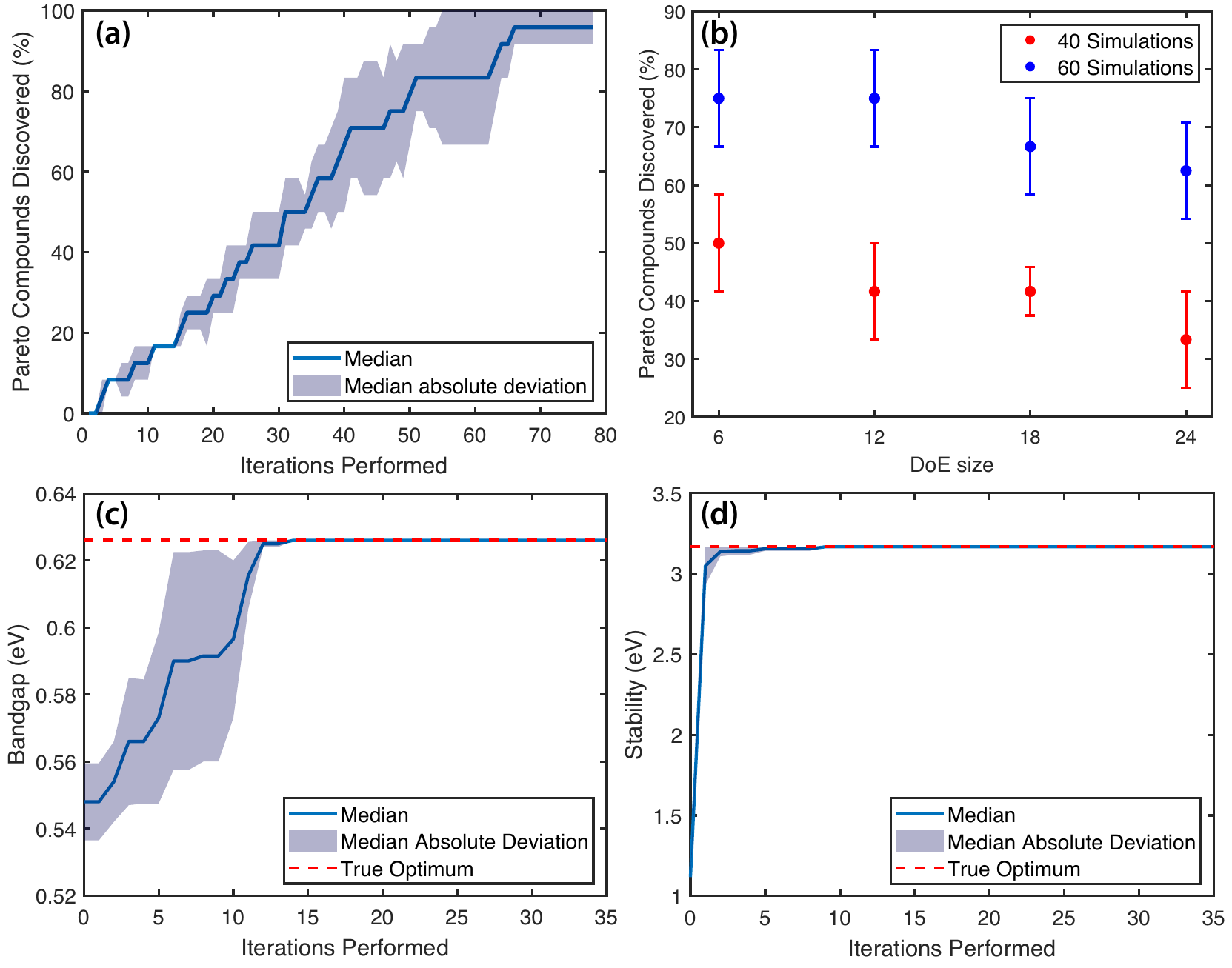}\vspace{-8pt}
  \caption{\sffamily\textbf{
  Robustness of the Adaptive Optimization Engine (AOE).}
  (a) The optimization history for 10 replicates of AOE, 
  each initialized with a distinct set of 12 initial DoE compounds. 
  In each trial, the initial DoE set consists of the same four known lacunar spinel compounds 
  and eight new compositions designed by the DoE procedure.
  %
  %
  Solid line shows the median percentage of true Pareto front compounds discovered at each iteration. The shaded area represents the median absolute deviation across 10 trials.
  \rw{(b) The fraction of Pareto front compounds discovered when the computational budget is fixed to 40 and 60 simulations. 
  Filled circles and their corresponding error bars represent the median and median absolute deviation respectively.}
  (c,d) The optimization history of 10 replicates of single-objective Bayesian optimization, targeting maximum band gap ($E_g$) and stability ($\Delta H^{*}_d$), respectively.
  The initialization method is the same as described in (a).
  %
  %
  Global optimum ($E^{*}_g = 0.626$\,eV, $\Delta H^{*}_d = 3.167$\,eV) is identified within 10\,\% exploration of design space.}
  \label{fig:fig6}
\end{figure}

\subsection{AOE performance}
\autoref{fig:fig4} (a) displays the results of the AOE. 
We successfully identify all 12 materials at the true Pareto front within 53 iterations (red asterisks, upper panel)---
compositions and objective-related properties are enumerated in \autoref{tab:table1}.
Combined with the 12 compounds from our initial DoE, 
we explored less than 25\% of the entire design space before identifying all lacunar spinels on the Pareto front.
Interestingly, Pareto-front compositions are mostly found with high EMI values, 
showing that our model makes beneficial recommendations 
on which composition to evaluate next.
High prediction uncertainty likely explains why a Pareto-front composition is not identified for some iterations with a large EMI. 
The EMI values reduce to nearly zero after all Pareto front compositions are identified (blue, upper panel) 
since all candidates not sampled are dominated by the Pareto front compounds.
We also show the absolute error in the LVGP-predicted $\Delta H_d$ (pink) 
and $E_g$ (orange) values of the evaluated composition at each iteration to further demonstrate the effectiveness of our model (\autoref{fig:fig4} (a)).
We find a general decreasing trend in error and therefore better model predictability as it becomes aware of more composition-property knowledge.
%

\autoref{fig:fig4} (b) shows the history of composition 
explored by the AOE for the first 60 iterations. 
%
The initial DoE sets are relatively scarcely distributed away from the \rw{true} Pareto front \rw{(marked as red asterisks)}, yet the model explores regions far from that covered by the DoE sets and is able to identify 75\% of Pareto front compositions within the first 40 iterations.
First, we begin to understand this performance by examining the distribution of elements sampled by the MOBO (\autoref{fig:fig4} (c)).
Our model does not exhibit much compositional bias upon sampling elements for the $A$ site; 
however, it shows clear preferences for choosing certain elements on other sites.
V and Mo are sampled more frequently on the basal M$^b$ site, while Nb and Ta are less favored on the apical M$^a$ site.
Se is also preferred over S and Te for the $Q$ site.
%

Then we examine the 2D latent space representations for both design objectives obtained after 60 iterations of AOE (\autoref{fig:fig4} (d) and (e)). 
The relative positioning of elements in the latent space reflects correlations in their influence on properties; 
elements in close proximity
exhibit similar impact. 
Interestingly, different transition metals exhibit distinct correlation patterns across various sites and objective properties. 
This variation leads us to conclude that 
($i$) the transition metals contribute to 
stability and band gap in different and unexpected ways, 
and ($ii$) the lack of any resemblance in element positioning in the site-dependent latent spaces, 
except for the M$^a$ site, 
to the periodic table indicates that chemical-intuition-based 
MIT design within the lacunar spinels is highly nontrivial. 
For example, chromium is located far from the other elements in the M$^a$ latent space, 
indicating that its influence on properties is distinct.
Indeed, Cr-containing compounds have significantly lower $E_g$ and higher $\Delta H_d$ (\autoref{fig:fig5}).

\begin{figure*}[tbh]
  \centering
  \vspace{3mm}
 \includegraphics[width=0.98\linewidth]{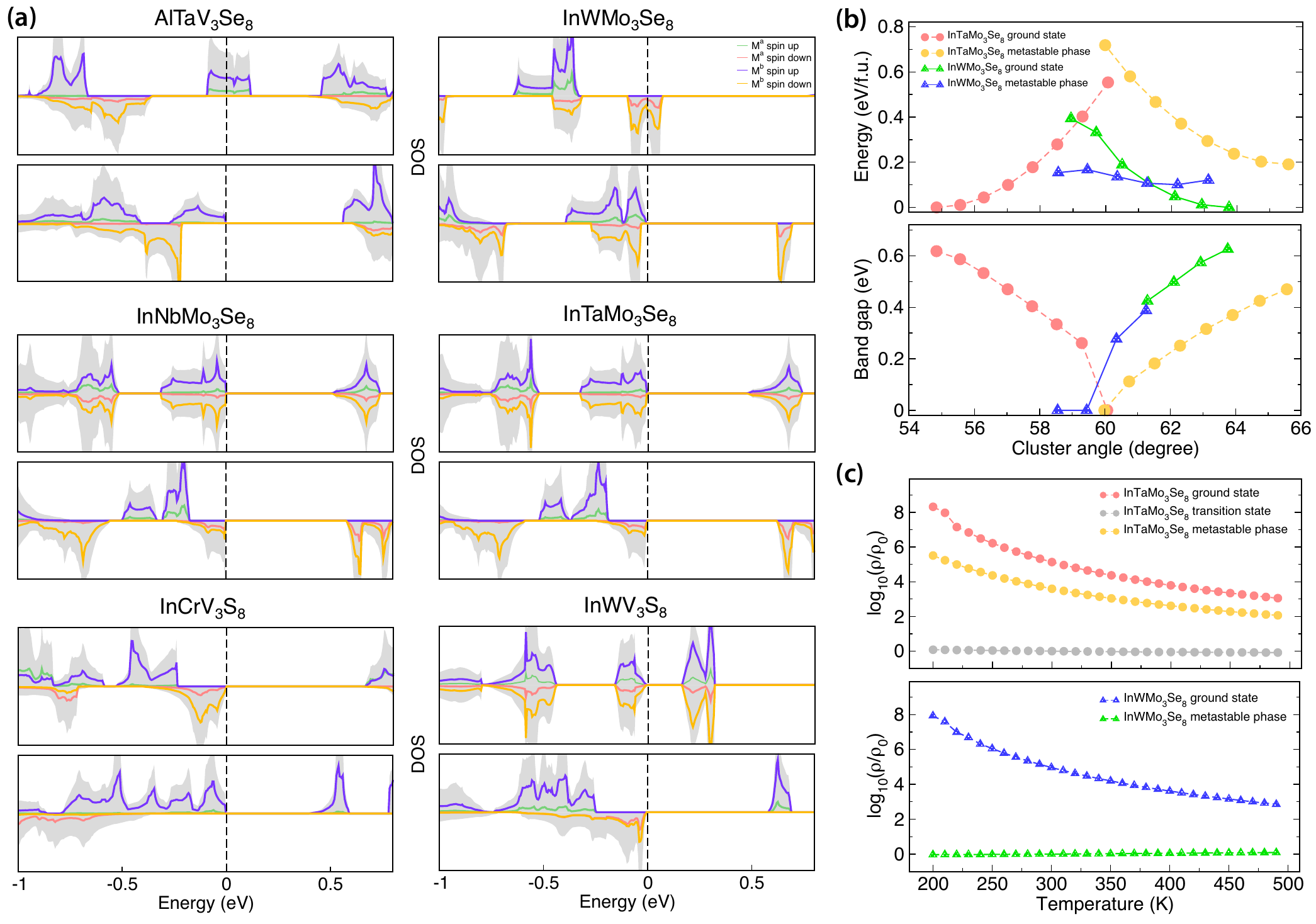}\vspace{-8pt}
  \caption{\sffamily 
  \textbf{DFT-simulated electronic properties of selected lacunar spinel compositions at the Pareto front.} 
  (a) The projected electronic density-of-states (DOS) of AlTaV$_3$Se$_8$, InWMo$_3$Se$_8$, InNbMo$_3$Se$_8$, InTaMo$_3$Se$_8$, InCrV$_3$S$_8$, and InWV$_3$S$_8$. 
  The lower panel of each composition shows the ground state electronic structure and the upper panel shows the DOS 
  of the metastable phase after the Jahn-Teller distortion.
  Both panels are normalized and span a range of 15 states per formula unit for each spin channel (vertical axis).
  AlTaV$_3$Se$_8$, InWMo$_3$Se$_8$ exhibit metal-insulator transitions whereas the other compounds show semiconductor-to-insulator transitions. 
  (b) The 
  DFT relative energies and band gaps of InWMo$_3$Se$_8$ and InTaMo$_3$Se$_8$ as a function of the cluster distortion angle $\theta_m$. 
  InTaMo$_3$Se$_8$ undergoes a semiconductor-to-insulator transition with a metallic intermediate state for  $\theta_m\approx60^{\circ}$.
  (c) Simulated DC resistivity of the compounds in (b) for their corresponding metallic, semiconducting, and intermediate states.}
  \label{fig:fig7}
\end{figure*}

The aforementioned performance is robust as revealed by our 
multi-trial results (\autoref{fig:fig6} (a)), where we find 
the AOE successfully identifies 90\,\% of the true Pareto-front compositions by exploring 30\,\% of the design space with different initial DoE sets. 
Since LHD is inherently random, repeating the DoE procedure
will lead to another randomly generated DoE set.
Therefore, we use this method to run multiple trials of AOE with different DoE sets. 
\ai{The size of DoE is another parameter for the designer to select in the AOE framework. Since the computational budget is often the bottleneck in discovery, the designer must allocate it wisely between the DoE and AOE. We investigated this problem using a set of four DoE sizes:  6, 12, 18, and 24, because there are six elements admissible at the M$^a$ site (\autoref{fig:fig6} (b)).
In each case, the  computational budget is fixed to 40 and 60 simulations and they 
are split between DoE size and AOE iterations. For example, 40 simulations can be split into DoE of size 6 and 34 iterations of AOE whereas a DoE of size 12 corresponds to  28 iterations of AOE, etc. 
Here, the four known gallium based compounds were not explicitly included in the DoE.
We find  that using a small DoE to initialize AOE (conversely, allocating more simulations to the AOE) is advisable, as its uncertainty guided exploration is more likely to discover Pareto compositions (\autoref{fig:fig6} (b)).}

Single-objective Bayesian optimization
on both band gap ($E_g$) and stability ($\Delta H_d$) are also performed 
using Expected Improvement acquisition criterion\cite{mockus1978application}, as shown in \autoref{fig:fig6} (b, c), respectively. 
Unsurprisingly, the model shows much higher efficiency in identifying the optimal composition than in the multi-objective task, where less than 10\,\% of the entire design space is explored.
We also notice that the model is always able to quickly infer the compound with highest stability, as depicted by the steep curve in \autoref{fig:fig6} (c).
Intuitively, thermodynamic stability is straightforward to linearize from  elemental reference states whereas the band gap is determined by the valence electronic structure and multiple interactions. Therefore, it might be easier for the model to decode the relationship between composition and stability, while learning the band gap dependency requires accumulating more knowledge.


\subsection{Pareto Compound Analysis}
We use DFT simulations to examine the properties of the identified Pareto-front compositions, focusing on 
$\Delta H_d$, $E_g$, and the Jahn-Teller active phonon  $\nu_\textrm{JT}$ involved in the MIT (\autoref{tab:table1}).
%
We find most Pareto-front compositions consist of two different cations on the M$^a$ and M$^b$ site, 
only three have $\mathrm{M}^a=\mathrm{M}^b$, with 75\,\% of the optimized materials being selenides.
GaV$_4$Se$_8$ is the only Pareto front compound previously synthesized, \rw{and verified to exhibit resistive-switching behavior under an applied electric pulse.}\cite{corraze2013electric}
All compounds exhibit $R3m$ symmetry 
and are dynamically stable in their ground state ($\nu_\textrm{JT}>0$). 
The phonon frequencies of the selenides, including $\nu_\textrm{JT}$ 
are lower than those of the sulfides. 
All of the designed lacunar spinels also exhibit semiconducting gaps with semilocal exchange-correlation and static Coulomb interactions and exhibit nonzero electric polarizations.
Compositions with larger band gaps tend to have lower stability as determined by $\Delta H_d$: 
2/3 are stable ($\Delta H_d>0$, indicating decomposition is endothermic), whereas four of the 12 compounds comprising Mo have small values of $\Delta H_d<0$, 
which could nonetheless be stable and synthesizable.\cite{Bartel2018,Aykol2018}
\rw{Typically, highly ionic materials with large electronic band gaps are also quite stable (e.g., NaCl). However, we find a clear  trade-off between these two properties for the Pareto front compositions. One possible reason is because all of these candidate materials are small-gap semiconductors (with $E_g < 0.65$\,eV) due to metal-metal and semiconvalent bonding while also being  polymorphous; therefore, these lacunar spinels are unlikely to follow the general trend. In addition, \autoref{fig:fig5} shows that the transition metals contribute to $E_g$ and $\Delta H_d$ in quite different ways, which could lead to this functionality-stability trade-off.
The AOE, however, does not posses knowledge of chemistry beyond the lacunar spinel family; yet, it is  able to resolve the $\Delta H_d$-$E_g$ relationship regardless of whether there is a trade-off or positive correlation.
These findings  reinforce the effectiveness of this model.}
%

%
Although the ground states of these materials are all semiconducting, 
we find two different electronic transitions upon traversing the ideal TMC geometry ($\theta_m=60^\circ$): 
the expected (Type I) metal-to-insulator transition
and an unexpected (Type II) semiconductor-to-insulator transition (SIT).
\autoref{fig:fig7} (a) shows the changes to the electronic structure for the MIT lacunar spinels 
AlTaV$_3$Se$_8$ and InWMo$_3$Se$_8$  
with the insulating state (lower panel) always lower in energy 
than the metastable metallic phase (upper panel) after the Jahn-Teller-type distortion 
($\theta_m\neq60^\circ$, \autoref{tab:table1}). 
%
%
The pDOS of these compounds show that the metallic state in the Type I transition 
arises from cluster distortion-triggered orbital ordering and occupancy changes, similar to the mechanism depicted in \autoref{fig:fig7} (b). 
However, the metallic states are different owing to the chemistry of the metals comprising the TMCs.
We also find that the basal M$^b$ site plays a more decisive role near the Fermi level 
with minor contribution from the apical M$^a$ site.
The M$^a$ site on the other hand, plays an active role in the Jahn-Teller-active phonon owing to differences in  atomic mass (\autoref{tab:table1}).
%
%
The remaining lacunar spinels in \autoref{fig:fig7} (a), InNbMo$_3$Se$_8$, InTaMo$_3$Se$_8$, InCrV$_3$S$_8$, and InWV$_3$S$_8$, exhibit a Type II transition.
The lower and upper panel show their ground and metastable state pDOS, respectively.
Interestingly, some compounds undergo singlet formation  and transform into a nonmagnetic 
phase (e.g., InNbMo$_3$Se$_8$) while others remain ferromagnetic after the cluster distortion (e.g.,  InCrV$_3$S$_8$) owing to 
competition between spin-pairing and magnetic interactions
\cite{khomskii2019molecules}. 
%
%

Last, we model the switching process and resistivity upon structural distortion for InWMo$_3$Se$_8$ (Type I) and  InTaMo$_3$Se$_8$ (Type II) by  modulating the amplitude of  the $\nu_\mathrm{JT}$ atomic displacements for each material in both the (insulating) ground and (metallic or semiconducting) metastable states.
%
The DFT-simulated energy and corresponding band gap at different cluster angles ($\theta_m$) are  shown in \autoref{fig:fig7} (b).
Both compounds show first-order transitions. 
Owing to the small changes in the TMC geometry required for switching, readily available external stimuli could be used to trigger the transitions\cite{vaju2008electric,camjayi2014first,Juraschek2017}.
%
The simulated DC resistivity of InWMo$_3$Se$_8$ and  InTaMo$_3$Se$_8$ 
clearly shows the promising functionality of these newly discovered compositions in the lacunar spinel family 
(\autoref{fig:fig7} (c)).
Since we successfully identify all 12 Pareto-front compositions by searching through less than 25\% of the design space,
our work demonstrates the efficiency of featureless adaptive materials discovery for electronic materials design.
The \rw{featureless} AOE is particularly useful when data availability and physical understanding of the target materials system is limited at either the atomic or microstructural scale.

\section{Discussion and outlook}
Our multiple property objectives of high stability and large insulating band gaps were
achieved  by using Bayesian optimization (BO) for MIT materials-composition design without explicitly constructing features (descriptors) via latent-variable Gaussian process implemented in our adaptive optimization engine.
We successfully identified 
all 12 Pareto-front lacunar spinel compositions by searching through less than 25\% of the design space.
\rw{Since the Utopian composition with both high functionality and stability (i.e., the upper right corner of \autoref{fig:fig4} (b)) cannot be realized, the Pareto front illustrates the trade-offs among objectives. This information is beneficial to materials scientist as it aids in the selection of candidate materials to further investigate or deploy.
The selection rules will depend on the designer’s preferences and whether to favor one property over others as well as their willingness to compromise. 
Therefore, we report the steps needed to identify all Pareto designs to quantify our model efficiency.}
Because these  materials have garnered much research attention in recent years owing to the richness of their fascinating physical behaviors (e.g., MITs, skyrmion lattices, and superconductivity),  we anticipate the newly identified lacunar spinels  will be pursued experimentally in search of these phenomena.
%
\begin{table}[t]
\begin{ruledtabular}
\centering
\caption{\label{tab:table1}\sffamily 
DFT-evaluated ground state properties of the Pareto front compounds. 
NOI is the number of iterations taken to discover the compound during the adaptive optimization process.
Values of $\Delta H_d>0$ (units of eV\,f.u.$^{-1}$) indicate  an endothermic reaction occurs and the stable compound disfavors decomposition. 
$E_g$ is the DFT band gap in eV.
$\nu_\textrm{JT}$ is the frequency (THz) of the Jahn-Teller-type phonon involving  the TMC.  
$P$ is the electric polarization in $\mu$C\,cm$^{-2}$. 
The value of $\theta_m$ in the insulating ground state 
and transition type, Type I (MIT) or Type II (SIT), are also specified.
}
\begin{tabular}{lccccccc}%
Compound & NOI & $\Delta H_d$ & $E_g$ & $\nu_\textrm{JT}$ & $P$ & $\theta_m$ & Type \\[0.2em] \hline
InWV$_3$S$_8$ & 4 & 0.09 & 0.58 & 5.83 & 0.56 & 65.0 & \RN{2}  \\
\hline
AlCrV$_3$Se$_8$ & 8 & 3.17 & 0.19 & 3.77 & 1.87 & 56.4 & \RN{2} \\
\hline
InMo$_4$Se$_8$ & 14 & -0.69 & 0.62 & 4.55 & 1.08 & 63.4 & \RN{1}\\
\hline
InWMo$_3$Se$_8$ & 19 & -0.99 & 0.63 & 4.43 & 0.24 & 63.8 & \RN{1}\\
\hline
InCrV$_3$S$_8$ & 20 & 2.59 & 0.40 & 4.75 & 0.28 & 56.6 & \RN{2}\\
\hline
AlCrV$_3$S$_8$ & 21 & 2.63 & 0.39 & 5.81 & 1.02 & 57.0 & \RN{2}\\
\hline
InCrV$_3$Se$_8$ & 25 & 3.10 & 0.22 & 3.45 & 0.58 & 56.0 & \RN{2}\\
\hline
InTaMo$_3$Se$_8$ & 28 & -0.88 & 0.62 & 4.25 & 1.38 & 54.8 & \RN{2}\\
\hline
AlTaV$_3$Se$_8$ & 38 & 0.56 & 0.56 & 3.90 & 0.15 & 57.3 & \RN{1}\\
\hline
AlV$_4$Se$_8$ & 47 & 1.06 & 0.46 & 4.08 & 2.80 & 54.9 & \RN{1}\\
\hline
InNbMo$_3$Se$_8$ & 49 & -0.66 & 0.59 & 4.44 & 0.75 & 55.2 & \RN{2}\\
\hline
GaV$_4$Se$_8$ & 53 & 1.18 & 0.44 & 4.09 & 2.37 & 55.0 & \RN{1}
\end{tabular}
\end{ruledtabular}
\end{table}

Although we have seen an increasing emphasis on using Bayesian optimization for materials design, previous work relied heavily upon handcrafted features, which is a challenging task, or single objective optimization. The former usually requires either knowledge of influential features based on theory and literature or large datasets to perform sensitivity analysis and correlation analysis to identify features that influence properties of interest. In the lacunar spinel MIT materials design, the scientific community is limited by chemical intuition as well as large datasets to identify appropriate features. This hinders the application of traditional BO implementations for MIT design. 
The propensity to use features arises mainly due to a lack of accurate and efficient machine learning methods to model categorical inputs. Here we showed
LVGP can circumvent feature identification by directly modelling elements as categorical variables. The mapping of the categorical variables into low-dimensional quantitative latent variables provides an inherent ordering for the categories and physics-based dimensionality reduction. Like conventional Gaussian process models, the LVGP model provides uncertainty quantification, which is crucial for employing the BO strategy for material composition optimization. LVGP enables featureless learning and subsequently featureless BO, making it a generic step forward in machine learning  and materials design. 

The AOE we demonstrated is theoretically more efficient than evolutionary algorithms for identifying the Pareto frontier in a complex, combinational design space. 
Although designing materials under a single criterion is more efficient, such efforts may not meet the requirements of deployment. 
For lacunar spinels investigated here, maximizing $E_g$ exclusively leads to an unstable composition while maximizing $\Delta H_d$ exclusively leads to a composition with a small bandgap. 
In contrast, MOBO identifies the Pareto front to delineate the trade-off between materials properties and allows the designer to choose compositions for detailed study. In this context, the need to perform more iterations of MOBO within the AOE is justified.
Indeed, it is typically not the sole goal to find all Pareto front designs, but rather to identify the best candidates within a limited research budget. The AOE clearly provides an efficient way to minimize the effort towards a better design by suggesting the next experimental design. 

\ai{Similar to forward materials design demonstrated here, inverse materials design\cite{Zunger2018}  can be cast as an optimization problem and tackled via the AOE framework. Although forward design is achieved with the objective of maximizing the desired properties, inverse design can be accomplished by redefining the objective as the minimization of the difference between the predicted and target properties. The design space, i.e., the choice of admissible elements, must be defined appropriately to ensure the target properties are achieved.}
To that end, our work advances materials innovation for forward and inverse 
design of both inorganic (as shown herein) and organic materials, such as identification of new quantum materials, design of protein sequence in biomaterials, and monomer sequence in polymeric materials.
It is particularly useful when data availability and physical understanding of the target materials system is limited at either the atomic or microstructural scale.
This methodology could be further extended  to mixed-variable optimization problems, e.g.,  co-design of composition and chemical stoichiometry through doping, which we are 
now actively developing.

\begin{acknowledgments}
The authors thank Dr.\ Danilo Puggioni at Northwestern University and 
Professors Ram Seshadri and Stephen Wilson at the University of California, Santa Barbara, for helpful discussions about this project.
This work was supported in part by the National Science Foundation (NSF) under award number DMR-1729303 and DMREF-1729473.
The information, data, or work presented herein was also funded in part by the Advanced Research Projects Agency-Energy (ARPA-E), U.S.\ Department of Energy, under Award Number DE-AR0001209.
The views and opinions of authors expressed herein do not necessarily state or reflect those of the United States Government or any agency thereof.
\textit{Ab initio} DFT simulations were performed on 
the DoD-HPCMP (Copper cluster) and Extreme Science and Engineering Discovery Environment (XSEDE), which is supported by NSF (ACI-1548562).

Y.W.\ and A.I.\ contributed equally to this work.
W.C.\ and J.M.R.\ conceived and administered the project.
Y.W.\ performed the DFT simulations and automated the AOE together with A.I.
A.I.\ developed the Design of Experiment procedure, implemented MOBO, and performed multiple trials of AOE.
Y.W.\ and A.I.\ wrote the first draft of the paper, which was revised based on input from all authors.

\end{acknowledgments}

\section*{Data Availability}
The data that support the findings of this study are available from the corresponding author upon reasonable request.

%



\appendix
\section{Density Functional Calculation Details}
%
%
We perform DFT simulations as implemented in 
the Vienna \textit{Ab initio} Simulation Package (VASP)\cite{kresse1996efficient, kresse1999ultrasoft}.
The projector augmented-wave (PAW) potentials\cite{blochl1994projector} 
are used for all elements in our calculations with the following  
valence electron configurations:
Al ($3s^{2}3p^{1}$),
Ga ($3d^{10}4s^{2}4p^{1}$), 
In ($4d^{10}5s^{2}5p^{1}$),
V ($3s^23p^63d^44s^1$), 
Nb ($4s^24p^64d^45s^1$), 
Ta ($5p^65d^46s^1$),
Cr ($3s^{2}3p^{6}3d^{5}4s^{1}$),
Mo ($4s^{2}4p^{6}4d^{5}5s^{1}$), 
W($5s^{2}5p^{6}5d^{5}6s^{1}$),
S ($3s^23p^4$), 
Se ($4s^24p^4$), 
and Te($5s^{2}5p^{4}$).
%
We use exchange-correlation potentials ($V_{xc}$) as implemented by Perdew-Burke-Ernzerhof (PBE)\cite{perdew1996jp}.
The effect of on-site Coulomb interactions (PBE$+U$) is considered with a $U$ value of 2.0\,eV for all 6 transition metals.
Previous studies have shown that such settings could nicely capture the complex electronic structures within the lacunar spinel family.\cite{wang2019assessing,kim2014spin} 
%
Numerous spin configurations are evaluated to ensure the global ground state is achieved and that those states are consistent with available experimental magnetic data.\cite{rastogi1983itinerant}
Spin-orbit interactions (SOI) are not considered in our calculations. %
Although  it has been shown that SOI leads to interesting molecular $j_{\mathrm{eff}}$ states,\cite{kim2014spin}
this order does not strongly affect the size of the ground state electronic band gaps, 
even $5d$ transition metals lacunar spinels.\cite{wang2019assessing}
A $\Gamma$-centered $6\times6\times 6$ $k$-point mesh with a 500\,eV kinetic energy cutoff is used.
We employ Gaussian smearing with a small 0.05\,eV width. 
For density-of-state calculations, we use the tetrahedron method 
with Bl\"{o}chl corrections.\cite{PhysRevB.49.16223}
Electric polarizations along the [111] direction are simulated using the Berry phase method\cite{resta1994macroscopic}.
%

The crystal structures of the existing lacunar spinels
are obtained from our previous DFT studies\cite{github_link_old},
structures of new compositions are obtained by replacing the elements on the corresponding crystallographic sites from existing structures.
We perform full lattice relaxations  
until the residual forces on each individual atom are less than 1.0\,meV\AA$^{-1}$.
The DFT-relaxed crystal structures of the Pareto front compositions are available at Ref.\  \onlinecite{github_link_new}.
We initialize the relaxation with various magnetic moment configurations, 
the converged configuration with the lowest energy is reported as the DFT ground state.
Zone center ($\mathbf{k}=\mathbf{0}$) phonon frequencies 
and eigendisplacements are obtained using the frozen-phonon method 
with pre- and post-processing performed with the Phonopy package\cite{phonopy}.
The decomposition pathways are automatically generated using Grand Canonical Linear Programming\cite{kirklin2013high} 
from the Open Quantum Materials Database\cite{kirklin2015open}.

Resistivity simulations are performed using electronic structures computed from VASP as previously described, but  with an increased $24\times24\times24$ $k$-point mesh
and the BoltzTrap2 package\cite{BoltzTraP2}. 
We also assume that all M$^a$ sites have the same orientation within the crystal.
In order to validate this model, we simulated a $2\times2\times2$ supercell of InNbMo$_3$Se$_8$ with one Nb atom oriented in a different direction from the other seven. We find that the ground state $E_g$ as well as $\Delta H_d$  exhibit negligible changes from the homogeneous description. We also compared the change in properties with the anti-ferromagnetic spin configuration using a doubled simulation cell with the ferromagnetic ground state. As before, we find there are no significant changes in the aforementioned properties.
These results are reasonable because the local structure of the TMC dictates the low-energy band structure near the Fermi level.\\ 
%


\section{Adaptive Optimization Engine (AOE) Implementation}
%
Conventional Gaussian process (GP) modelling has been developed for only quantitative design variables 
and the associated correlation functions cannot handle categorical variables. 
To overcome this limitation, 
LVGP maps 
each categorical variable to a 2D Cartesian latent space\cite{zhang2019latent,zhang2020bayesian}, 
establishing a numerical representation 
for different categories.
With this mapping, the covariance model over categorical design variables can be any standard GP covariance model for quantitative variables, e.g., the Gaussian correlation function.
In the AOE, two independent LVGP models with Gaussian correlation function are fit at each iteration 
to predict $E_g$ and $\Delta H_d$, respectively. 
In each LVGP model, categorical variables $A$, M$^a$, M$^b$ and $Q$ 
are represented by a 2D numerical latent variable vector to evaluate their correlation. 
Note that each categorical variable resides in its unique latent space.
For the LVGP model predicting $E_g$, 
let $\boldsymbol{z}^{A}=[z^{A}_{1},z^{A}_{2}]$ denote the latent variable for the $A$ site. 
Similarly, $\boldsymbol{z}^{\mathrm{M}^a}$, $\boldsymbol{z}^{\mathrm{M}^b}$, and $\boldsymbol{z}^{Q}$ 
denote the latent variables for M$^a$, M$^b$ and $Q$ site, respectively. 
Then, the Gaussian correlation ($\rho$) between $E_g$ of two compounds, 
e.g. GaMoV$_3$S$_8$ and AlNbW$_3$Se$_8$, is:
\begin{multline}
\rho \left( E_g^{\mathrm{GaMoV}_3\mathrm{S}_8}, E_g^{\mathrm{AlNbW}_3\mathrm{Se}_8} \right) = \\
\exp \left( 
-\|\boldsymbol{z}^{\mathrm{Ga}} - 
\boldsymbol{z}^{\mathrm{Al}}\|_{2}^{2}
-\|\boldsymbol{z}^{\mathrm{Mo}} - 
\boldsymbol{z}^{\mathrm{Nb}}\|_{2}^{2} \right.\\
\left.
-\|\boldsymbol{z}^{\mathrm{V}} - 
\boldsymbol{z}^{\mathrm{W}}\|_{2}^{2}
-\|\boldsymbol{z}^{\mathrm{S}} - 
\boldsymbol{z}^{\mathrm{Se}}\|_{2}^{2}  
\right)
\end{multline}
%
where $\|.\|_2$ represents the Euclidean 2-norm. This procedure is used to compute the correlation matrix for properties of all evaluated compositions. 
%
The positioning of latent variables $\boldsymbol{z}^{A}$,
$\boldsymbol{z}^{\mathrm{M}^a}$, $\boldsymbol{z}^{\mathrm{M}^b}$, and $\boldsymbol{z}^{Q}$ in their corresponding latent space are estimated via MLE as described in Ref \onlinecite{zhang2019latent}.
The LVGP model for $\Delta H_d$ also utilizes the 2D latent variable representation $\boldsymbol{\kappa}^{A}$, $\boldsymbol{\kappa}^{\mathrm{M}^a}$, $\boldsymbol{\kappa}^{\mathrm{M}^b}$, and $\boldsymbol{\kappa}^{Q}$ 
as previously defined to evaluate the correlation 
$\rho ( \Delta H_d^{\mathrm{GaMoV}_3\mathrm{S}_8}, \Delta H_d^{\mathrm{AlNbW}_3\mathrm{Se}_8} )$ in a similar manner.

Multiobjective Bayesian optimization includes first 
considering the lacunar spinel family 
$A$M$^{a}$M$^{b}_3Q_8$ 
with $A \in \{\mathrm{Al,Ga,In}\}$, 
$\mathrm{M}^a \in \{\mathrm{V, Nb, Ta, Cr, Mo, W}\}$, $\mathrm{M}^b \in \{\mathrm{V, Nb, Ta, Mo, W}\}$ 
and $Q \in \{\mathrm{S,Se,Te}\}$. 
The design space ($\boldsymbol{C}$) comprises 270 compounds, 
each compound is represented by four design variables $A, \mathrm{M}^a, \mathrm{M}^b \textrm{ and } Q$ 
with three, six, five, and three choices, respectively.
Our objective is to maximize $E_g$ and $\Delta H_d$, 
which is represented in standard optimization formulation as:
\begin{equation}
\min_{\boldsymbol{c} \in\, \boldsymbol{C}}
{-E_g(\boldsymbol{c}), -\Delta H_d(\boldsymbol{c})}\,.
\end{equation}
Starting from the initial dataset, the AOE evaluates new candidate compounds 
by gauging their improvement in the design objectives.
Here, we use the expected maximin improvement (EMI) metric\cite{bautista2009sequential} to guide the adaptive sampling framework. 
The Maximin Improvement ($I_M$) for compound $\boldsymbol{c}$ is: 
\begin{multline}
I_M(\boldsymbol{c}) = \\
\min_{\boldsymbol{c_i} \in\, \boldsymbol{C}_{PF}}
\left\{
\max\left(\widetilde{E_g}(\boldsymbol{c}) - \widetilde{E_g}(\boldsymbol{c_i}),
\widetilde{\Delta H_d}(\boldsymbol{c}) - \widetilde{\Delta H_d}(\boldsymbol{c_i}),0\right)
\right\}
\label{eq:eq3}
\end{multline}
where $\boldsymbol{C}_{PF}$ is the current set of Pareto front compositions. 
To facilitate the comparison in \autoref{eq:eq3}, 
we scale the value of each design objective $P$ using the scheme 
$\widetilde{P}(\cdot)= (P(\cdot)-P^{min}) / (P^{max}-P^{min})$ where $P^{max} \textrm{ and } P^{min}$ are the maximum and minimum value of property observed so far. 
\rw{By scaling the properties, we ensure all design objectives are comparable and viewed equally.}
The EMI of compound $\boldsymbol{c}$ is defined as the expected value of $I_M$:
\begin{equation}
\mathrm{EMI}(\boldsymbol{c}) = \mathbb{E}[I_M(\boldsymbol{c})]\,.
\end{equation}
We evaluate the EMI through Monte Carlo sampling with 500 trials.
%
%
At each AOE iteration, the EMI is calculated for all compositions that are not yet present in the data repository. 
The composition with largest EMI will be sampled next in property evaluation and then added to the data repository.\\

\bibliography{reference}
\vspace{1em}

\end{document}